\documentclass[twocolumn, prx, superscriptaddress]{revtex4-2}

\usepackage[utf8]{inputenc}
\usepackage[english]{babel}
\usepackage[T1]{fontenc}
\usepackage{amsmath}
\usepackage{hyperref}
\usepackage{xcolor}
\hypersetup{
    colorlinks,
    linkcolor={red!50!black},
    citecolor={blue!50!black},
    urlcolor={blue!80!black}
}

\usepackage[capitalise]{cleveref}
\usepackage{gensymb}
\usepackage{tikz}
\usepackage{lipsum}
\usepackage{bm}

\usepackage{multirow}

\newcommand{\acr}[1]{{STEP}}

\begin{document}

\title{Inferring pointwise diffusion properties of single trajectories with deep learning}

\author{Borja Requena}
\affiliation{ICFO -- Institut de Ci\`encies Fot\`oniques, The Barcelona Institute of Science and Technology, Av. Carl Friedrich Gauss 3, 08860 Castelldefels (Barcelona), Spain}

\author{Sergi Mas\'o}
\affiliation{Facultat de Ci\`encies, Tecnologia i Enginyeries, Universitat de Vic -- Universitat Central de Catalunya (UVic-UCC), C. de la Laura,13, 08500 Vic, Spain}

\author{Joan Bertran}
\affiliation{Facultat de Ci\`encies, Tecnologia i Enginyeries, Universitat de Vic -- Universitat Central de Catalunya (UVic-UCC), C. de la Laura,13, 08500 Vic, Spain}

\author{Maciej Lewenstein}
\affiliation{ICFO -- Institut de Ci\`encies Fot\`oniques, The Barcelona Institute of Science and Technology, Av. Carl Friedrich Gauss 3, 08860 Castelldefels (Barcelona), Spain}
\affiliation{ICREA, Pg. Llu\'is Companys 23, 08010 Barcelona, Spain}

\author{Carlo Manzo}
\affiliation{Facultat de Ci\`encies, Tecnologia i Enginyeries, Universitat de Vic -- Universitat Central de Catalunya (UVic-UCC), C. de la Laura,13, 08500 Vic, Spain}

\author{Gorka Mu\~noz-Gil}%
\email{munoz.gil.gorka@gmail.com}
\affiliation{Institute for Theoretical Physics, University of Innsbruck, Technikerstr. 21a, A-6020 Innsbruck, Austria}

\begin{abstract}
In order to characterize the mechanisms governing the diffusion of particles in biological scenarios, it is essential to accurately determine their diffusive properties.
To do so, we propose a machine learning method to characterize diffusion processes with time-dependent properties at the experimental time resolution.
Our approach operates at the single-trajectory level predicting the properties of interest, such as the diffusion coefficient or the anomalous diffusion exponent, at every time step of the trajectory.
In this way, changes in the diffusive properties occurring along the trajectory emerge naturally in the prediction, and thus allow the characterization without any prior knowledge or assumption about the system.
We first benchmark the method on synthetic trajectories simulated under several conditions.
We show that our approach can successfully characterize both abrupt and continuous changes in the diffusion coefficient or the anomalous diffusion exponent. 
Finally, we leverage the method to analyze experiments of single-molecule diffusion of two membrane proteins in living cells: the pathogen-recognition receptor DC-SIGN and the integrin $\alpha5\beta1$. The analysis allows us to characterize physical parameters and diffusive states with unprecedented accuracy, shedding new light on the underlying mechanisms.
\end{abstract}

\maketitle

\section{Introduction}
Advances in optical imaging have made it possible to observe single molecules in living biological systems~\cite{mockl2014super}. When combined with particle tracking algorithms, these techniques allow tracing the movement of individual molecules, viruses, and organelles with nanometric precision, enabling the study of transport mechanisms in complex biological environments. Through the biophysical characterization of trajectories, we can extract meaningful parameters to describe physical and biological processes. However, accurately quantifying the trajectories remains a challenging task due to the stochastic nature of these processes and to experimental drawbacks, such as imaging noise and the emitter's photophysics~\cite{manzo2015review}.

\begin{figure*}
    \centering\includegraphics[width=0.8\textwidth]{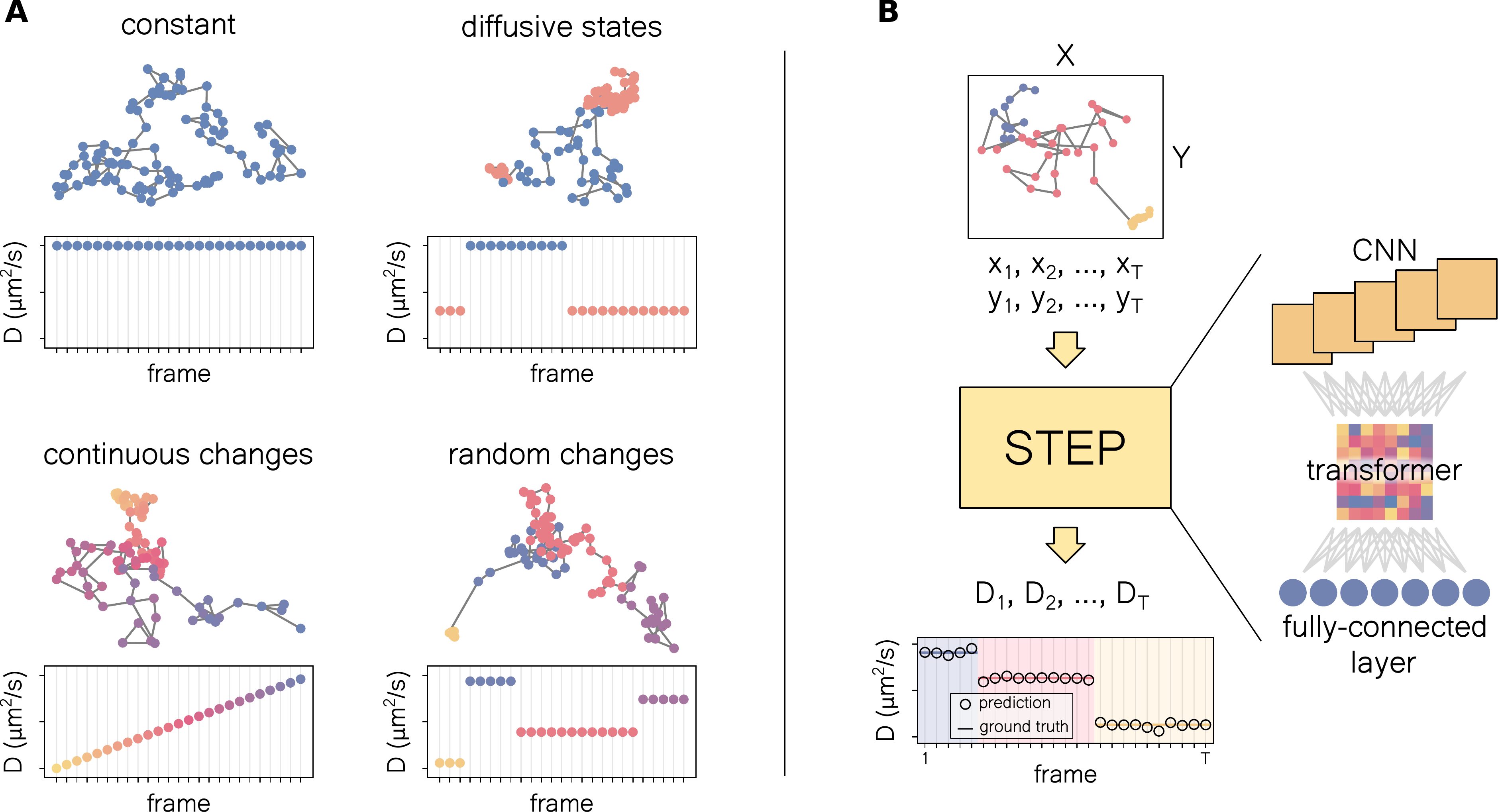}
    \caption{\textbf{Heterogeneous trajectories and the \acr{} pipeline.}
    (\textbf{A}) Examples of trajectories and the corresponding diffusion coefficient $D$ as a function of time for: constant $D$; changes within a discrete set of states with fixed $D$; continuous and monotonous change of $D$; switch between random $D$s.
    (\textbf{B}) Schematic of the pipeline of \acr{}: an input trajectory is fed to the architecture, which consists of a stack of CNN, a transformer encoder, and a pointwise feedforward layer. The model's output is the pointwise prediction of the diffusion parameter of interest (in this case $D$).
    }
    \label{fig:scheme}
\end{figure*}

In cellular systems, a widespread diffusion feature is the occurrence of time-dependent changes of motion~\cite{yin2018detection}. Typically, these changes are associated with transient interactions with other components~\cite{saha2015diffusion,bag2015plasma,low2011erbb1} resulting in the sudden variation of a parameter, e.g., the diffusion coefficient, which can switch between a discrete~\cite{sabri2020elucidating}, or continuous set of levels~\cite{jeon2016protein, lampo2017cytoplasmic, manzo2015weak}. Furthermore, they can induce smooth changes such as those associated with the spatiotemporal heterogeneity of the environment~\cite{jeon2014scaled}. Examples of trajectories undergoing this kind of diffusion are schematically depicted in \cref{fig:scheme}A.

Trajectories with time-dependent diffusion properties pose an additional challenge to characterize the motion of individual particles, which has been tackled with different approaches.
For trajectories displaying abrupt changes, segmentation methods~\cite{vega2018multistep, arts2019particle, lanoiselee2017unraveling, yin2018detection} are a valuable strategy but cannot deal with long-range correlations and often offer limited time resolution due to temporal averaging.
On the other hand, model-dependent methods such as the hidden Markov model (HMM) have been quite successful in describing heterogeneous diffusion~\cite{bronson2009learning, persson2013extracting, monnier2015inferring}, although they require prior knowledge about the diffusive states involved and their kinetic scheme. Recently, data-driven approaches have shown remarkable capabilities to extract information from individual stochastic trajectories, even in the presence of changes of diffusion properties~\cite{munoz2021objective, verdier2022variational,  pineda2023geometric}. 

In this work, we propose \acr{}, a  method based on state-of-the-art deep-learning architectures to extract pointwise diffusion features from individual trajectories without any prior information (see \cref{fig:scheme}B).
\acr{} features the most recent advances in sequence-to-sequence learning~\cite{Sutskever2014seq2seq}, which have shown impressive results in natural language processing tasks and beyond~\cite{brown2020language, reed2022generalist, taylor2022galactica}.
It combines convolutional~\cite{lecun1989backpropagation} and attention layers~\cite{vaswani2017attention} to cope with the presence of short and long-range correlations,  providing remarkable performance over trajectories of any length and in the presence of noise.

The article is structured as follows: first, we show the ability of \acr{} to predict diffusion properties, such as the diffusion coefficient and the anomalous diffusion exponent, on simulated data reproducing experimentally relevant scenarios. Then, we analyze simulated trajectories with smoothly varying diffusion coefficients, showing that \acr{} correctly finds the expected scaling. Finally, we use \acr{} to study two experimental data sets obtained by the tracking of single-molecule live-cell imaging experiments and reporting the motion of two membrane receptors: \emph{i}) the pathogen-recognition receptor DC-SIGN, which has been associated with random changes of diffusion coefficients~\cite{manzo2015weak}; and \emph{ii}) the $\alpha 5 \beta 1$ integrin, expected to be transiently arrested by binding to the cytoskeleton and the extracellular matrix~\cite{kanchanawong2022organization}. 

\section{Results}
\label{sec:results}

\subsection*{The \acr{} architecture}
Recently, we have witnessed an enormous effort in the development of deep-learning approaches to study diffusive processes~\cite{munoz2021objective}.
Previous works usually focused on characterizing diffusive properties of single trajectories, i. e., predicting an overall or average diffusive parameter for each input trajectory~\cite{munoz2020single,granik2019single, Bo2019RNN,kowalek2019classification, Seckler2022BayesianDL}.
Recent works have proven the suitability of this approach to study complex phenomena in different experimental scenarios~\cite{jamali2021anomalous, munoz2022stochastic}.
With \acr{}, we propose a sequence-to-sequence approach that translates position coordinates into diffusion properties providing their pointwise prediction at every time step of the input trajectory~\cite{pineda2023geometric}.
\acr{} translates an input trajectory of arbitrary length $T$ into a sequence of $T$ elements containing the diffusive properties of interest, as illustrated in~\cref{fig:scheme}B.
While it is effectively impossible to characterize diffusion from a single displacement, \acr{} uses the whole trajectory as context to perform the prediction at every point.

This approach allows us to study trajectories where diffusion properties can vary over time with different patterns: from trajectories with constant diffusive properties to trajectories that switch between discrete diffusive states or where diffusive parameters change continuously over time (see \cref{fig:scheme}A for examples).
Unlike previous works, where expert input is needed in order to choose an appropriate method, \acr{} can be seamlessly applied to any diffusive data. Importantly, it does not rely on prior assumptions, such as the number of changepoints~\cite{gentili2021characterization, argun2021classification} or the properties of the expected diffusive states~\cite{Maizon2021DL}.

State-of-the-art architectures for diffusion characterization rely on very different approaches~\cite{Verdier2021GNN,manzo2021extreme, Li2021WADNet}. Several models obtain outstanding performance by combining recurrent neural networks (RNNs) that account for long-range correlations~\cite{Bo2019RNN,argun2021classification} with convolutional neural networks (CNNs) focusing on local features~\cite{garibo2021efficient,munoz2022stochastic,firbas2023characterization}.
In \acr{}, we use a similar scheme but replace the RNN with a transformer encoder~\cite{vaswani2017attention}, the cutting-edge architecture for sequence modeling, and we stack XResNet blocks~\cite{he2019bag} to build the convolutional part  (see Methods). Local features are extracted through a CNN and fed to the transformer, thus processing local and long-range features simultaneously (see~\cref{fig:scheme}B). 
Finally, we use a pointwise fully-connected layer of non-linear neurons to obtain the appropriate output dimension.
We provide a detailed description of the method in~\cref{app:ml_architecture}.

\begin{figure*}
    \centering\includegraphics[width=\textwidth]{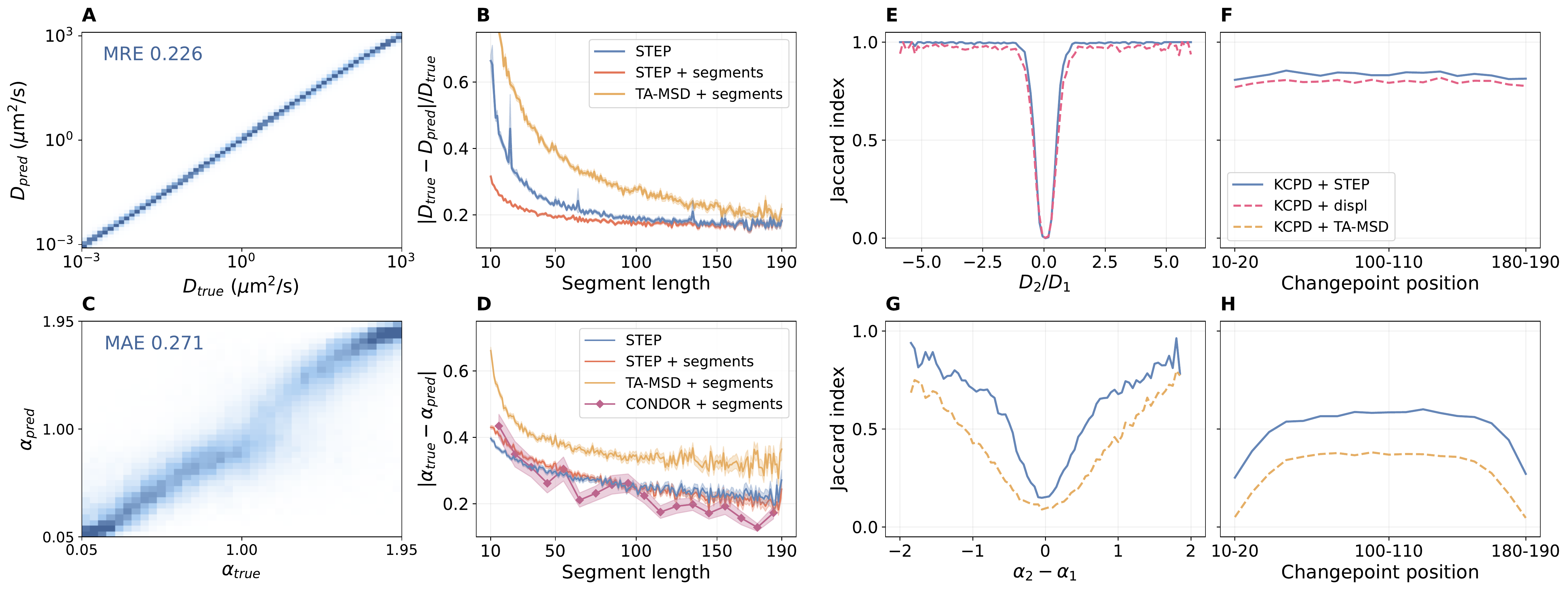}
    \caption{
    \textbf{Time-dependent diffusion properties prediction.}
    (\textbf{A}) 2D histogram of the predicted diffusion coefficient $D$ compared to the ground truth. The mean relative error (MRE) over the whole test set is 0.226.
    (\textbf{B}) Relative error for the prediction of $D$ as a function of the segment length.
    (\textbf{C}) 2D histogram of the predicted anomalous diffusion exponent $\alpha$ of FBM segments compared to the ground truth. The MAE over the test set with all the diffusion models is 0.271. 
    (\textbf{D}) Mean absolute error (MAE)  for the prediction of $\alpha$ as function of the segment length.
    (\textbf{E} to  \textbf{H}) Jaccard index for the changepoint detection problem as a function of:
    (\textbf{E})  the ratio between consecutive segment $D$s ,
    (\textbf{F}) the changepoint position for $D$, 
    (\textbf{G}) the difference between consecutive segment $\alpha$s,
    and (\textbf{H}) the changepoint position for $\alpha$. 
    }
    \label{fig:benchmark}
\end{figure*}

\subsection*{Pointwise prediction of diffusion properties}
\label{sec:benchmark_pointwise}

We first validate \acr{} on the task of inferring the diffusion coefficient pointwise from simulated trajectories reproducing transient Brownian motion with abrupt changes of diffusion coefficient. The diffusion coefficient can randomly vary in the range $D\in[10^{-3}, 10^3]$ and the dwell time in each diffusion coefficient is drawn from an exponential distribution. Further details of the simulations and the data sets are described in~\cref{app:datasets}. A 2D histogram of the ground truth versus the predicted diffusion coefficient shows that \acr{} can precisely determine the diffusion coefficient on the whole range of $D$ values included in the simulations (\cref{fig:benchmark}A) with an overall relative error $|D_{\text{true}}-D_{\text{pred}}|/D_{\text{true}} = 0.226$.

To further explore the performance of \acr{}, we calculate the relative error as a function of the segment length (i.e. the dwell time for each diffusion coefficient) that expectedly shows an improvement at longer length (blue line in \cref{fig:benchmark}B).
As a baseline to compare \acr{}'s precision in estimating the diffusion coefficient, we use the linear fit of the time-averaged mean squared displacement (TA-MSD) (see~\cref{app:diffusion}). However, while \acr{} can obtain pointwise predictions of $D$ from heterogeneous trajectories, the TA-MSD needs trajectories of constant $D$ and was thus fed with pre-segmented trajectories. In spite of this disadvantage, \acr{} provides better results (blue vs yellow line in \cref{fig:benchmark}B). Notably, when \acr{} is provided with pre-segmented trajectories, we observe a further improvement, with a nearly 2-fold reduction of the error at short segment length (red line in \cref{fig:benchmark}B), demonstrating outstanding prediction capabilities.

We then examine the ability of \acr{} to predict the anomalous diffusion exponent $\alpha$. We consider heterogeneous trajectories simulated according to different underlying models (further details are provided in \cref{app:diffusion}) composed of segments with $\alpha$ suddenly changing in the range $\alpha \in [0.05,2]$. The 2D histogram of the ground truth vs the predicted $\alpha$ for trajectories undergoing fractional Brownian motion (FBM) (\cref{fig:benchmark}C) shows that \acr{} successfully predicts the anomalous diffusion exponent with a mean absolute error $| \alpha_{\rm true}-\alpha_{\rm pred}| = 0.271$, obtaining results in line with the top-scoring approaches for this task~\cite{argun2021classification,garibo2021efficient} in the Anomalous Diffusion challenge~\cite{munoz2021objective}.

In~\cref{fig:benchmark}D, we further report the mean absolute error for the inference of $\alpha$ as a function of the segment length and in comparison to the results obtained through the linear fit of the TA-MSD in logarithmic space and CONDOR~\cite{gentili2021characterization}, the best-in-class approach in the corresponding task of the AnDi Challenge~\cite{munoz2021objective}. As discussed earlier, both methods were provided with pre-segmented trajectories, whereas \acr{} deals directly with heterogeneous trajectories. \acr{} strongly outperforms the TA-MSD approach and shows performance comparable to CONDOR. For this task, providing pre-segmented data to \acr{} marginally improves its performance for long segments, whereas it even reduced it for short ones. This result suggests that segment length is more important than the exact knowledge of the segment edges and \acr{} effectively combines local and global information. 

In~\cref{app:complementary_D}, we extend the assessment of the performance of \acr{} as a function of the localization precision and the dwell-time duration/number of segments of the trajectories, showing that the method can be efficiently applied in a wide range of experimental conditions. Additionally, in~\cref{app:alpha_by_models}, we extend the analysis about anomalous diffusion showing the performance of~\acr{} as a function of the underlying diffusion models and the localization precision.

\subsection*{Detecting diffusive changepoints in heterogeneous trajectories}
\label{sec:benchmark_accuracy}

For trajectories undergoing sudden changes of diffusion properties, the exact knowledge of the points at which these properties' changes occur is crucial to infer temporal properties and kinetic rates of the system and fully characterizing the underlying physical process.
While \acr{} does not explicitly detect changepoints, its output provides a precise estimation of the diffusion property which is supposed to change, hence simplifying the task of changepoint detection and location with respect to the use of raw data. 
To highlight this capability, we compare the results obtained by a state-of-the-art kernel changepoint detection (KCPD) method~\cite{Celisse2018kcp,Arlot2019kcp} when applied to \acr{}'s predictions and to the timetrace of trajectory displacements.
To assess the performance of the method, we compute the Jaccard index (JI) considering as a true positive any changepoint predictions lying within a threshold distance $\mathcal{E}$ from the corresponding ground truth.

We first quantify the performance of the method to detect changes of the diffusion coefficient in heterogeneous trajectories performing Brownian motion. We use a benchmark data set with trajectories exhibiting a single changepoint with $\mathcal{E}=5$.
Through the application of the KCPD algorithm on the prediction of \acr{}, we can successfully detect the changepoints with high accuracy even for consecutive segments whose diffusion coefficients are just one order of magnitude apart, as we show in~\cref{fig:benchmark}E (blue line).
Furthermore, the method is robust with respect to the changepoint position within the trace (\cref{fig:benchmark}F (blue line)).
In contrast, when we apply KCPD directly over the trajectory displacements (dashed purple lines) we observe a decrease in performance over the whole range of diffusion coefficient ratio.
On average, \acr{} produces a 20\% reduction in error reaching an average JI of 0.833 compared to 0.796 obtained with the raw displacements.

We perform a similar analysis to detect changes in the anomalous diffusion exponent in FBM trajectories.
Since the anomalous diffusion exponent is an asymptotic property and cannot be easily calculated from the raw data, to build our baseline, we compute $\alpha$ with a linear fit of the TA-MSD on a log-log scale using a sliding window of 30 time steps, which we then feed into the KCPD algorithm (dashed yellow lines).
Expectedly, the larger the differences between segment parameters, the better we can detect the changepoints, as we show in~\cref{fig:benchmark}G. We obtain a 30\% reduction of error by using \acr{} with respect to the baseline method, achieving an average JI of 0.515 and 0.297, respectively, with $\mathcal{E}=20$.
However, the plot of the performance shows that finding changes in $\alpha$ is a harder task than finding changes in $D$. Moreover, we also observe a performance drop when the changepoints are near the trajectory edges (\cref{fig:benchmark}H). 
In these cases, we deal with short segments whose anomalous diffusion exponent can be hard to determine, as they rely on the arising of long-range correlations.

\subsection*{Revealing continuous changes of diffusion properties}
\label{sec:continuous_changes}

When considering heterogeneous trajectories in the biological context, the typical behavior one expects is represented by particles undergoing diffusion with piecewise constant properties that can suddenly change, e.g., as the result of specific interactions with other biological components.
However, the presence of molecular crowding and gradients of concentration can produce a continuous variation of diffusion properties over time. These changes might be challenging to detect due to the limited spatiotemporal resolution of the experiments or the lack of specific approaches for trajectory analysis. 
Since \acr{} predicts pointwise diffusion properties in a model-free fashion, it inherently features the capability to perform this kind of analysis, even without dedicated training.

\begin{figure}
    \centering
    \includegraphics[width=0.8\columnwidth]{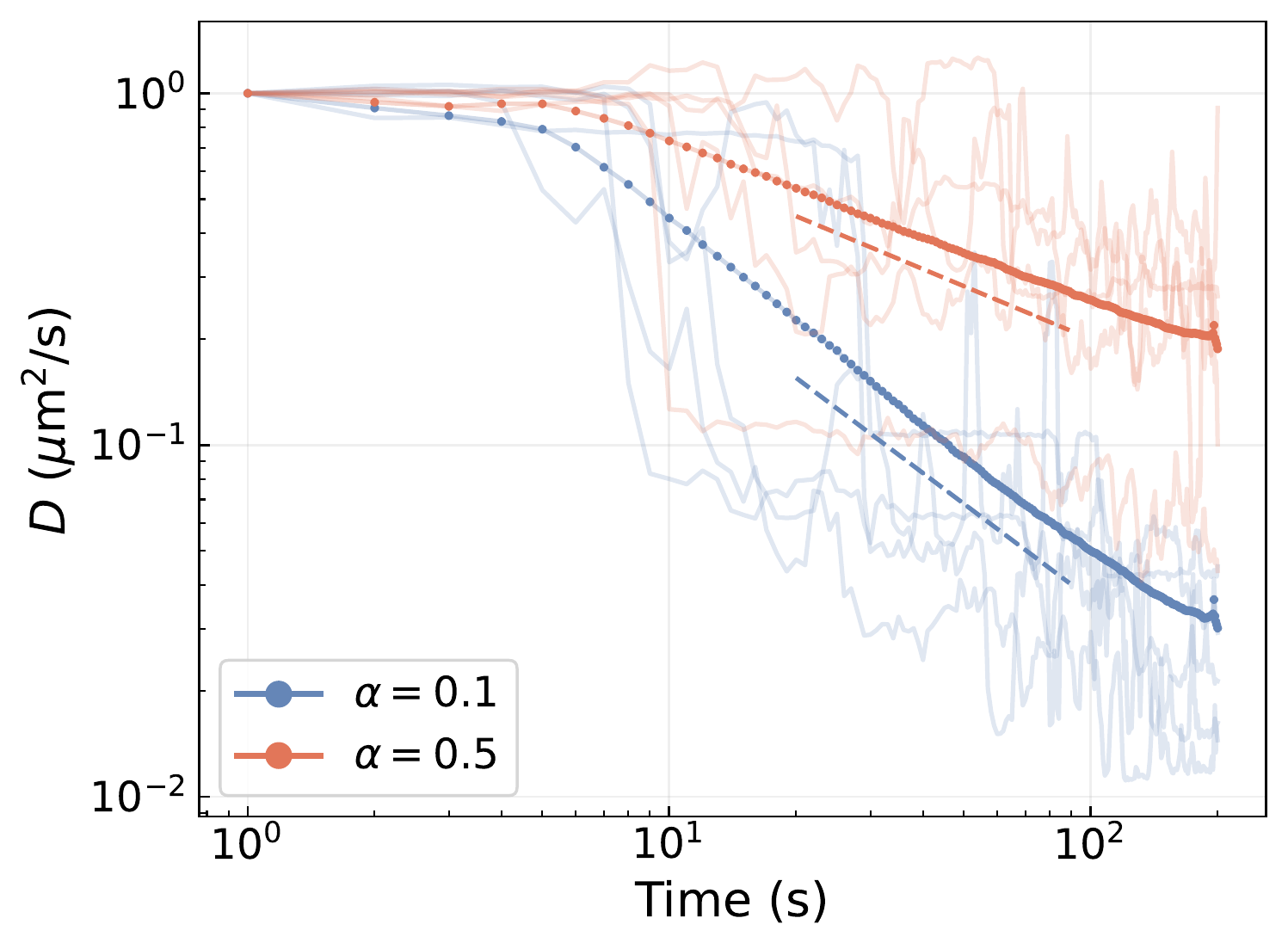}
    \caption{\textbf{Continuous changes of diffusion properties.} Predictions of the diffusion coefficient of two sets of SBM trajectories with $\alpha = 0.1$ (blue) and $0.5$ (red). The thin continuous lines correspond to predictions for exemplary single trajectories, bold continuous lines show the average over 3000 trajectories, and the dashed lines show the theoretically expected scaling for every set.}
    \label{fig:sbm}
\end{figure}

\begin{figure*}
    \centering
    \includegraphics[width=\textwidth]{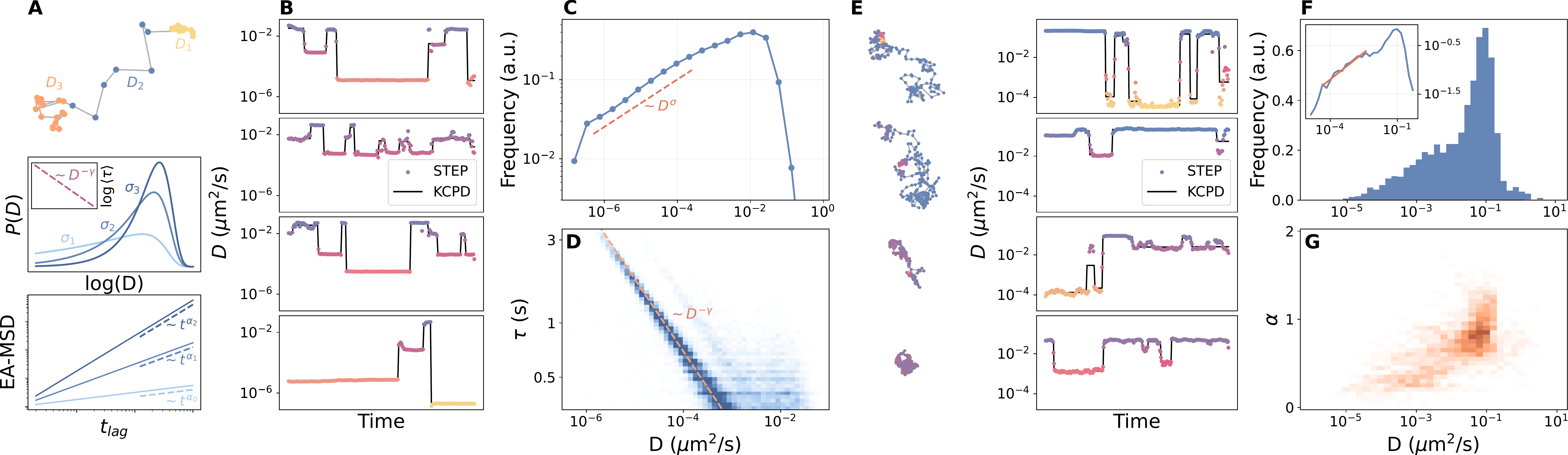}
    \caption{\textbf{Switch between random diffusive states of the pathogen-recognition receptor DC-SIGN.} 
    (\textbf{A}) Characteristic features of the ATTM model: an exemplary trajectory undergoing changes of diffusion coefficient, the distribution of diffusion coefficient, and the relation between the diffusion coefficient and the dwell time.
    (\textbf{B}) Predictions of the diffusion coefficient obtained by applying \acr{} to simulated ATTM trajectories (dots) and the result of applying the changepoint analysis (black line).
    (\textbf{C}) Distribution of $D$ obtained through the analysis described in (\textbf{B}),  showing the expected power law behavior at small $D$.
    (\textbf{D}) Relation between $D$ and the dwell time $\tau$ obtained through the analysis described in (\textbf{B}), showing the expected power law behavior.
    (\textbf{E}) Examples of experimental trajectories of DC-SIGN with the corresponding predictions obtained for $D$ (dots) and the changepoint analysis (black line).
    (\textbf{F}) Histogram of the distribution of $D$ obtained for the experimental trajectories. Inset: power-law fit at small $D$.
    (\textbf{G}) 2D histogram of $D$ and $\alpha$ obtained for the experimental trajectories.
}
    \label{fig:exp_attm}
\end{figure*}

To evaluate the performance of \acr{} on smoothly-varying trajectories, we rely on simulations of scaled Brownian motion (SBM)~\cite{jeon2014scaled}. SBM trajectories are characterized by a time-dependent diffusion coefficient with a power-law relationship $D(t)\sim t^{\alpha-1}$, where $\alpha$ is the anomalous diffusion exponent. Further details about simulations are given in~\cref{app:datasets}.

In \cref{fig:sbm}, we show the predictions obtained for the diffusion coefficient at every time step of trajectories with $\alpha = 0.1$ and $0.5$. The shaded lines represent the predictions obtained for individual trajectories which, despite the fluctuations, already indicate the decreasing trend. Averaging over 3000 trajectories with the same $\alpha$ (bold lines) reveals the correct scaling as compared with the expected power law (dashed lines).
Furthermore, the inference of $\alpha$ correctly provides a nearly constant value throughout the trajectory, as expected (see~\cref{app:alpha_by_models} for additional details), and points towards a low false positive rate with respect to changes in anomalous diffusion.

\subsection*{Characterizing anomalous diffusion from changes of normal diffusive properties}
\label{sec:attm_experiments}

To test the potential of \acr{} for the analysis of experimental trajectories, we use it to study the motion of the pathogen-recognition receptor DC-SIGN expressed in Chinese hamster ovarian cells~\cite{manzo2012neck}.
Previous analysis of these experiments revealed the occurrence of anomalous diffusion and weak ergodicity breaking as a consequence of changes of diffusion coefficient~\cite{manzo2015weak}. This behavior was described in the framework of the annealed transit-time model (ATTM)~\cite{massignan2014nonergodic}, whose main features are schematically summarized in \cref{fig:exp_attm}A.

In brief, ATTM depicts Brownian diffusion randomly switching diffusion coefficient $D$, with values sampled from a distribution with a power-law behavior $D^{\sigma-1}$ for small $D$ and a fast decay for $D \rightarrow  \infty$.
The model further assumes a correlation between $D$ and the dwell time $\tau$ of the form $\tau(D)\sim D^{-\gamma}$ to predict an anomalous diffusion exponent $\alpha = \sigma / \gamma$.
Therefore, the correct characterization of both the distribution of $D$ and $\tau$ is crucial to corroborate the compatibility with the underlying model. In the original work, changes of diffusivity were detected through a changepoint analysis~\cite{montiel2006quantitative} but the sensitivity and the time-resolution of the method did not allow a thorough investigation of this behavior.

To demonstrate that \acr{} enables better characterization of these data, we first use simulated ATTM trajectories.
We segment the trajectories applying the KCPD algorithm introduced in the previous sections over the \acr{} predictions, as we show in ~\cref{fig:exp_attm}B. Thus, we assign to each segment a single $D$, taking the average segment prediction, and a $\tau$. We successfully recover the power-law behavior of $D$ (\cref{fig:exp_attm}C) and the power-law relationship between $\tau$ and $D$ (\cref{fig:exp_attm}D).

Then, we apply this approach to the DC-SIGN trajectories of Ref.~\cite{manzo2015weak}. The results confirm the occurrence of diffusivity changes between segments of nearly-constant diffusion coefficient and with variable duration, as we show in~\cref{fig:exp_attm}E. Interestingly, our approach reveals twice as many changepoints as the previous analysis.

\begin{figure*}
	\centering
	\includegraphics[width=\textwidth]{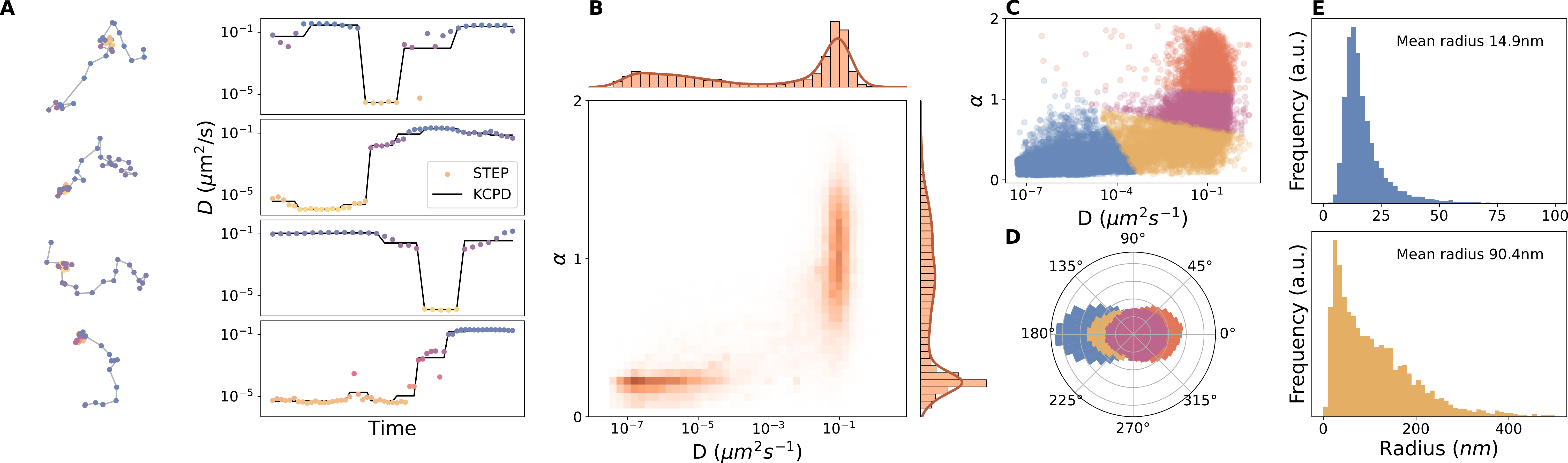}
	\caption{\textbf{Multi-state diffusion of the integrin $\bm{\alpha5\beta1}$.} (\textbf{A}) Examples of experimental trajectories of the integrin $\alpha$5$\beta$1 with the corresponding predictions obtained for $D$ (dots) and the changepoint analysis (black line). (\textbf{B}) 2D histogram of $D$ and $\alpha$ with the respective marginal distributions. (\textbf{C}) Scatter plot of the predictions obtained for $D$ and $\alpha$ at the segment level, color-coded according to a clustering analysis performed with a \emph{k}-means algorithm. (\textbf{D}) Distribution of the turning angle for the four clusters of segments obtained as in (\textbf{C}). (\textbf{E}) Distribution of the confinement radius for the clusters showing restrained diffusion.}
	\label{fig:exp_integrin}
\end{figure*}

The distribution of $D$ obtained for trajectory segments spans several orders of magnitude, as we show in the histogram of~\cref{fig:exp_attm}F.
For small $D$, it displays a behavior compatible with a power-law with exponent $\sigma\approx 0.37$ over nearly three decades (inset of \cref{fig:exp_attm}F), compatible with the ATTM. Notably, this behavior could not be directly verified in the original article.
In principle, our method would allow us to verify the correlation between $D$ and dwell time, as we have shown in the simulations. However, this task is limited by the variable trajectory length~\cite{tsunoyama2018super} and by the lack of statistics, in particular for segments at small $D$.  

As a further test, we predict the anomalous diffusion exponent with \acr{}. We assign a single $\alpha$ by taking the average prediction of each segment. The results reported in ~\cref{fig:exp_attm}G show an interesting correlation between $D$ and $\alpha$ that suggests a more complex diffusion pattern, involving the occurrence of anomalous diffusion also at the level of individual segments.

\subsection*{Characterizing multi-state diffusion processes}
\label{sec:integrin_experiments}

We use \acr{} to analyze experimental trajectories of the integrin $\alpha$5$\beta$1 diffusing in the membrane of HeLa cells. Integrins are transmembrane receptors for the extracellular matrix (ECM) in focal adhesions, which mechanically link the ECM and actin filaments in the cytoplasm and activate signaling pathways involved in cell migration, proliferation, or apoptosis~\cite{kanchanawong2022organization}. The dynamics of the integrin $\alpha$5$\beta$1 is influenced by interactions with fibronectin and actin-binding proteins~\cite{rossier2012integrins, tsunoyama2018super}. Its motion has been reported to switch from fast free-diffusion to slow free-diffusion and immobilization, as well as exhibiting rearward actin-driven movement.

We use \acr{} to predict both the diffusion coefficient and the anomalous diffusion exponent for the integrin $\alpha$5$\beta$1 trajectories. Then, we segment the trajectories by applying the KCPD method to both predictions at once. In this way, we assign every segment a unique $D$ and $\alpha$ by taking the average prediction over the segment. Examples of the results are shown in~\cref{fig:exp_integrin}A and the joint distribution of $D$ and $\alpha$ in~\cref{fig:exp_integrin}B. The visual inspection of \cref{fig:exp_integrin}B reveals two main clusters centered around ($D = 10^{-6} \mu$m$^2$/s, $\alpha = 0.25$) and  ($D = 0.1 \, \mu$m$^2$/s, $\alpha = 1$).
The 2D histogram of the same parameters calculated at the pointwise level (pre-segmentation) does not show any major differences with respect to \cref{fig:exp_integrin}B.

However, the unsupervised clustering of the data using a \emph{k}-means algorithm in combination with model selection performed with the elbow method~\cite{satopaa2011finding} reveals four clusters of segments (\cref{fig:exp_integrin}C) characterized by different motion features.
The first two clusters show a rather restrained motion, with integrins spending 40\% of the time in a state characterized by $D = 1.2\time10^{-5} \mu$m$^2$/s and $\alpha=0.23$, and 14\% of the time with $D = 0.06 \,\mu$m$^2$/s and $\alpha = 0.46$.
For both clusters, the distribution of angles between successive steps shows a peak centered at 180$\degree$, indicating backward movements due to reflection at potential boundaries, as we show in~\cref{fig:exp_integrin}D.
The confinement radius of the first cluster has a median of $14.9\,$nm (st. dev. $\pm 13.3\,$nm), which is comparable to the localization precision of these experiments. This allows us to associate it with protein immobilization.
The second cluster shows confined motion within areas with a broad distribution of sizes, as we see in~\cref{fig:exp_integrin}E, and a median radius of $90.4\,$nm (st. dev. $\pm 98.2\,$nm).
The third cluster represents the 29\% of the total recording and shows minor deviations from Brownian motion with $\alpha = 0.88$ and a nearly uniform angle distribution, and has an average $D = 0.10 \, \mu$m$^2$/s, close to the value typically reported for this protein.
Interestingly, the analysis pinpoints a fourth population, corresponding to a 20\% of the total recording, undergoing superdiffusion with $\alpha = 1.3$ and $D = 0.14 \, \mu$m$^2$/s, and with a persistent direction of motion between consecutive steps (\cref{fig:exp_integrin}D).

\section{Discussion}

In this work we present \acr{}, a machine learning method to predict diffusion properties from individual trajectories at every time step.
The method relies on a combination of state-of-the-art machine learning architectures that take into account correlations at different time scales.
The presented approach is especially appealing to analyze trajectories from particles undergoing heterogeneous motion, where changes in diffusion properties occur over time. 
Moreover, it does not require prior knowledge of the underlying physical process or the temporal resolution at which changes in diffusion occur.

To illustrate the predictive power of \acr{}, we benchmark it on simulated trajectories in various conditions. We show its ability to predict piecewise constant diffusion properties, such as the diffusion coefficient or the anomalous diffusion exponent, in noisy and short trajectories.
Furthermore, we demonstrate that \acr{} boosts up the accuracy of a changepoint detection algorithm to detect the time at which diffusion changes take place. 
Importantly, we also prove the suitability of our method to study continuous changes of diffusion.
To further showcase the potential applications of the method, we study trajectories obtained by tracking live-cell single-molecule imaging experiments of proteins of the plasma membrane.
First, we characterize the motion of the pathogen recognition receptor DC-SIGN, which was shown to exhibit random changes in the diffusion coefficient. Our analysis confirms such a hypothesis and improves the accuracy with which we detect these changes. Moreover, our results suggest the occurrence of more complex phenomena that need further investigation. 
Secondly, we study the diffusion of the integrin $\alpha_5\beta_1$. In agreement with previous works, our analysis confirms the existence of different diffusion modes and allows their precise classification according to the diffusion coefficient, the anomalous diffusion exponent, and the levels of spatial constraint.

We believe that \acr{} represents a first step towards a new class of machine learning algorithms to study dynamic systems through a sequence-to-sequence approach. Focusing on the instantaneous prediction of the property of interest  enables the characterization of the trajectories at experimental time resolution without averaging and filtering and minimizes the prior knowledge needed to perform the analysis. As such, the information retrieved with \acr{} can provide information about diffusion properties with unprecedented resolution and thus shed light on the underlying physical processes of a variety of systems.

\section*{Materials and Methods}
\label{sec:methods}

\subsection*{Machine learning model}

In diffusion phenomena, we deal with complex statistical signals that can exhibit various types of time correlations. Therefore, we need a machine learning (ML) model that can capture correlations at different time scales.
Furthermore, we often encounter trajectories with very different lengths, even in the same experiment. Hence, it is crucial that the model is \textit{length-independent} to ensure that it is as applicable as possible.
In this work, we propose to use a ML technique based on the advances of sequence-to-sequence learning~\cite{Sutskever2014seq2seq}, which has shown impressive results in natural language processing tasks and beyond~\cite{Brown2020gpt3, Alayrac2022flamingo, Reed2022gato}.
Analogously to a translation task, we implement a model that takes a particle trajectory as input and outputs the diffusion properties of interest at each time step. In this way, the input and the output have the same length. 

In our framework, the input trajectory, $\mathbf{x}$, is a $d$-dimensional vector of arbitrary length $T$ whose elements, $x_t$, correspond to the particle position at each time step, $t\in[1, T]$. In this work, we consider two-dimensional trajectories ($d=2$), but the method is easily adapted to any dimension.
On the other hand, the output is a one-dimensional vector of length $T$ whose elements contain the value of a diffusion property of interest. For instance, if we study the behavior of the diffusion coefficient along a trajectory, the output elements would be the diffusion coefficient at each time step, $D_t$. 
We refer the reader to~\cref{fig:scheme} for a schematic representation of the procedure.
In case we are interested in more than one property, we can either extend the output dimension, e.g., predicting $[D_t, \alpha_t]$, or implement an independent model for each one of them. In this work, we have opted for the second option. 

We propose an architecture combining convolutional and self-attention mechanisms. Interestingly, very recent works have shown analogous strategies, both with supervised~\cite{pineda2023geometric} and unsupervised approaches~\cite{kabbech2022identification}.
First, the input trajectory is processed by a series of convolutional layers that capture the short-range correlations. We implement them following the XResNet architecture~\cite{he2019bag}.
Then, the result follows through a transformer encoder~\cite{vaswani2017attention}, which can capture global correlations. Finally, we use pointwise fully-connected layers to reach the desired output dimension.

We refer to~\cref{app:ml_pipeline} for a detailed description of the architecture and its training procedure. We also provide a full library containing the code and detailed explanations on how to reproduce the results of this work in Ref.~\cite{Requena2022step}.

\subsection*{Cell culture and plating}
For the live-cell single-molecule imaging experiments involving the integrin ${\rm \alpha5\beta1}$, HeLa cells were cultured in DMEM (Gibco, 11960-044), supplemented with 10\% (v/v) fetal bovine serum (FBS, Sigma). Cells were tested for mycoplasma contamination using PCR (Biotools kit, 4542).
For fluorescence imaging, glass-bottom dishes (IBIDI, 81158) were coated with fibronectin (FN, Sigma, F2008) by placing 10$\,\mu$g/mL FN on the glass for 1$\,$h at 37$\,\degree$C, and then blocked with BSA 2$\,$mg/mL for 1$\,$h at 37$\,\degree$C. Cells were plated at a density of 5$\times$10$^4\,$cells/dish and cultured for 24$\,$h prior to use.

\subsection*{Preparation of half-antibody fragments}
Half-antibody fragments were obtained following a protocol similar to the one used in~\cite{low2011erbb1}. Briefly, mouse anti-human integrin ${\rm \alpha5}$ antibody (50$\,\mu$L; BD Biosciences, 610633) was dialyzed (ThermoFisher, Slide-A-Lyze MINI Dialysis Device, 2K) against PBS overnight at room temperature to replace the commercial buffer. Then, antibodies were reduced with 1$\,$mM DTT for 30$\,$min at room temperature and dialyzed again, using Slide-A-Lyze MINI Dialysis Device 2K, for 4$\,$h at room temperature against PBS to remove DTT. To avoid reassociation of reduced antibodies, sulfhydryl groups were blocked by incubating with iodoacetamide 20$\,$mM for 1$\,$h at 4$\,\degree$C with agitation. Iodoacetamide was then removed from the reaction by dialysis overnight at 4$\,\degree$C. Finally, reduced antibodies were biotinylated with a 10-fold molar excess of EZ-Link Sulfo-NHS\_LC\_Biotin (Thermo Scientific) for 30 min at room temperature with agitation and stored at 4$\,\degree$C until use.

\subsection*{Single-molecule labeling}
Biotinylated half-antibody fragments were conjugated to streptavidin-coated quantum dots (QD655 streptavidin conjugate, Invitrogen, Q10123mp). Cells were washed 3 times with washing buffer (PBS with 6\% BSA) and labeled with half antibody-quantum dots (about 1$\,$nM) in washing buffer (200$\,\mu$L per dish) for 15$\,$min at 37$\,\degree$C, followed by two washes. 

\subsection*{Live-cell single-molecule imaging}
Imaging was performed using a Leica DMi8 fluorescence microscope. Samples were illuminated in total internal reflection fluorescence (TIRF) geometry. Excitation was achieved with a CW laser (Obis, Coherent, $\lambda$=488$\,$nm, $<$1$\,$kW/cm$^2$). Fluorescence was recorded using an oil-immersion objective (Leica, 100X, NA=1.47) and an sCMOS camera (Photometrics 95B) with appropriate filters (Chroma). Movies were recorded at a frame rate of 33$\,$Hz. A microscope environment chamber (Okolab) was used to keep cells in a 5\% CO$_2$ atmosphere while recording.

\subsection*{Single-particle tracking}
Particle detection and tracking were performed using u-track~\cite{jaqaman2008robust}. The detection (Gaussian Mixture-Model Fitting) and tracking parameters were optimized based on visual inspection and performance diagnostic of the resulting detection and tracking. 
All image and data analysis tasks were performed in MATLAB 2020a and more recent versions (The MathWorks, Natick, MA). Videos  were loaded into MATLAB using Bio-Formats~\cite{linkert2010metadata}.

\bibliographystyle{unsrt}
\bibliography{biblio}

\section*{Acknowledgments}

The authors acknowledge Montserrat Masoliver-Prieto and Marta Cullell-Dalmau for their valuable help with the experimental procedures.
GMG acknowledges support from the European Union (ERC, QuantAI, Project No. 10105529) and the Austrian Science Fund (FWF) through the SFB BeyondC F7102.
CM  acknowledges support through grant RYC-2015-17896 funded by MCIN/AEI/10.13039/501100011033 and ``ESF Investing in your future'', grants BFU2017-85693-R and PID2021-125386NB-I00 funded by MCIN/AEI/10.13039/501100011033/ and FEDER ``ERDF A way of making Europe'', and grant AGAUR 2017SGR940 funded by the Generalitat de Catalunya.
BR and ML acknowledge support from: ERC AdG NOQIA; Ministerio de Ciencia y Innovaci\'on Agencia Estatal de Investigaciones (PGC2018-097027-B-I00/10.13039/501100011033, CEX2019-000910-S/10.13039/501100011033, Plan Nacional FIDEUA PID2019-106901GB-I00, FPI, QUANTERA MAQS PCI2019-111828-2, QUANTERA DYNAMITE PCI2022-132919, Proyectos de I+D+I “Retos Colaboraci\'on” QUSPIN RTC2019-007196-7); MICIIN with funding from European Union NextGenerationEU(PRTR-C17.I1) and by Generalitat de Catalunya; Fundaci\'o Cellex; Fundaci\'o Mir-Puig; Generalitat de Catalunya (European Social Fund FEDER and CERCA program, AGAUR Grant No. 2021 SGR 01452, QuantumCAT \textbackslash $\,$ U16-011424, co-funded by ERDF Operational Program of Catalonia 2014-2020); Barcelona Supercomputing Center MareNostrum (FI-2022-1-0042); EU Horizon 2020 FET-OPEN OPTOlogic (Grant No 899794); EU Horizon Europe Program (Grant Agreement 101080086 — NeQST), National Science Centre, Poland (Symfonia Grant No. 2016/20/W/ST4/00314); ICFO Internal “QuantumGaudi” project; European Union’s Horizon 2020 research and innovation program under the Marie-Skłodowska-Curie grant agreement No 101029393 (STREDCH) and No 847648 (“La Caixa” Junior Leaders fellowships ID100010434: LCF/BQ/PI19/11690013, LCF/BQ/PI20/11760031, LCF/BQ/PR20/11770012, LCF/BQ/PR21/11840013).
Views and opinions expressed in this work are, however, those of the author(s) only and do not necessarily reflect those of the European Union, European Climate, Infrastructure and Environment Executive Agency (CINEA), nor any other granting authority. Neither the European Union nor any granting authority can be held responsible for them

\appendix

\section{Diffusion properties}
\label{app:diffusion}
In this section, we briefly highlight some of the main characteristics of normal and anomalous diffusion. We refer the reader to Refs.~\cite{metzler2014anomalous, klafter2011first} for a nice and thorough introduction to the field.

Diffusion trajectories are often described by means of their mean squared displacement (MSD)  which, in the case of Brownian motion, shows a linear scaling with time, i.e. MSD $\propto D t$, where $D$ is the diffusion coefficient. However, there can be deviations from such linear scaling, resulting in a power-law relation between the MSD and time, i.e. MSD $\propto K_\alpha t^\alpha$, where $\alpha$ is defined as the anomalous diffusion exponent and $K_\alpha$ is an effective diffusion coefficient. The former allows us to distinguish between normal (or Brownian) diffusion ($\alpha = 1$) and anomalous diffusion ($\alpha \neq 1$).

The appearance of anomalous diffusion can be associated with very different phenomena, from the arising of correlations in the motion of the diffusing particle to the presence of spatiotemporal heterogeneity. To account for most of these phenomena, we follow Ref.~\cite{munoz2021objective} and consider five anomalous diffusion models with specific ranges for the anomalous diffusion exponent: annealed transient time model (ATTM)~\cite{massignan2014nonergodic} with $\alpha\in[0.05, 1]$, continuous-time random walk (CTRW)~\cite{Scher1975CTRW} with $\alpha\in[0.05, 1]$, fractional Brownian motion (FBM)~\cite{Mandelbrot1968FBM} with $\alpha\in[0.05, 1.95]$, L\'evy walk (LW)~\cite{Klafter1994LW} with $\alpha\in[1.05, 2]$, and scaled Brownian motion (SBM)~\cite{Lim2002SBM} with $\alpha\in[0.05, 2]$.

\section{Machine learning pipeline}
\label{app:ml_pipeline}

Here, we provide a detailed explanation of the machine learning approach followed to obtain the results described throughout this work.

As we briefly mention in the main text, we train two different models: one for the diffusion coefficient task, and one for the anomalous diffusion exponent. 
We report the results regarding the prediction of the diffusion coefficient and anomalous diffusion exponent in~\cref{sec:results}.
We implement both models following the same principles with very minor differences. In this section, we describe the architecture and the training process that we follow and, when needed, highlight the differences between models.

We provide the source code with extended explanations on how to reproduce the results in~\cite{Requena2022step}. We make extensive use of the \emph{PyTorch}~\cite{pytorch} and \emph{fastai}~\cite{fastai} libraries to implement the architecture and the training procedure. The kernel changepoint detection method (KCPD) was implemented using the \emph{ruptures} Python library~\cite{truong2020selective}.

\subsection{Architecture details}
\label{app:ml_architecture}
\begin{figure}
    \centering
    \includegraphics[width=\columnwidth]{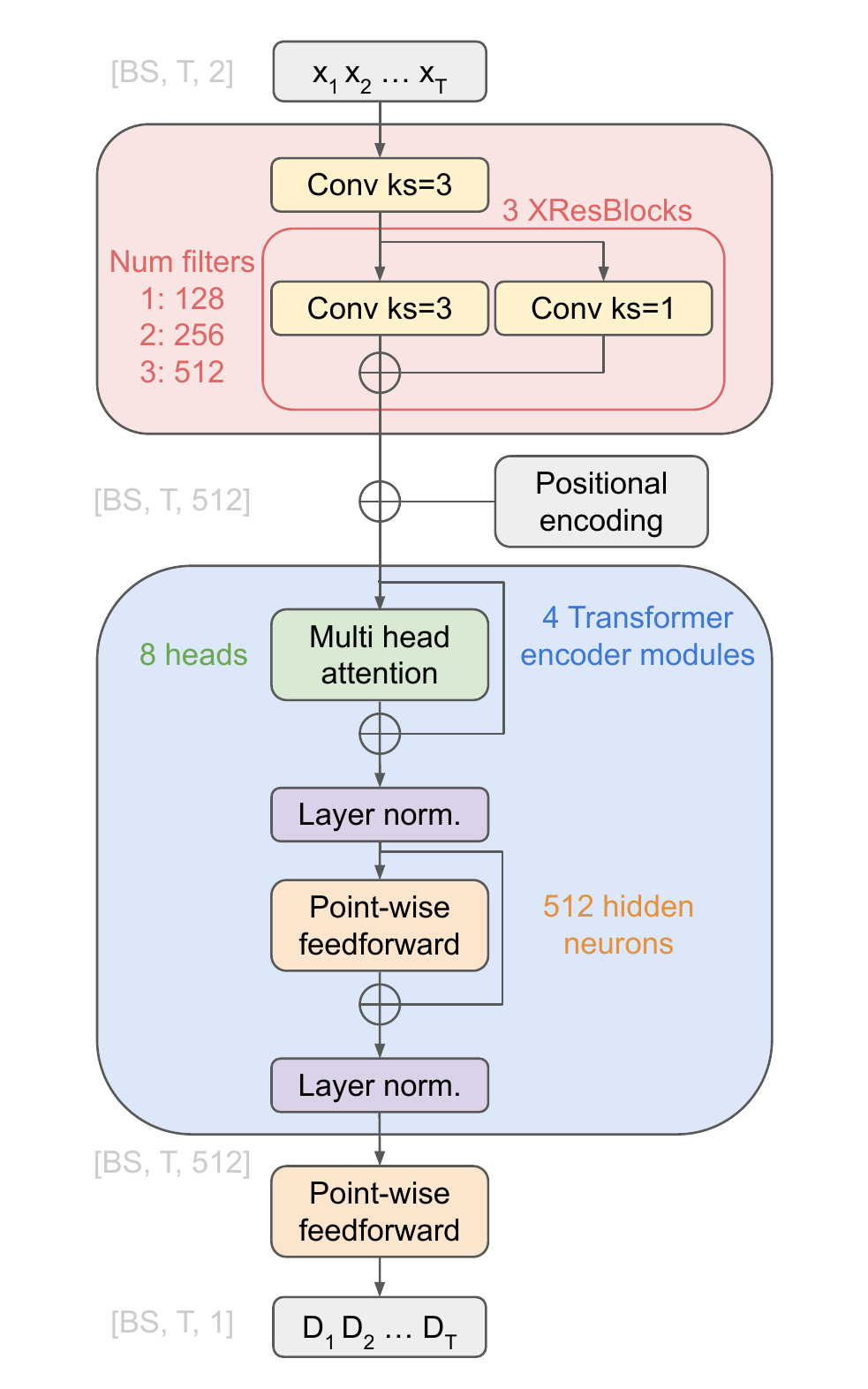}
    \caption{Machine learning architecture representation}
    \label{fig:ml_model}
\end{figure}

We propose to use a model that takes a trajectory $\mathbf{x}$ as input and outputs the target diffusion properties at each time step. The input trajectory is a $d$-dimensional vector of arbitrary length $T$, whose elements, $x_t$, correspond to the particle position at every time step $t$.
Then, the output is a one-dimensional vector of length $T$, whose elements correspond to the diffusion property of interest at every time step, e.g., $D_t$ in the case of the diffusion coefficient.
See~\cref{fig:ml_model} for further details about the dimensions. Throughout this work, we mainly consider trajectories of dimension $d=2$.

The model we propose consists of three main modules: an initial convolutional part that processes the input trajectory; a self-attention-based part that feeds on the features extracted by the previous one; a shallow pointwise fully connected feedforward module that provides the desired output dimensions. The entire architecture is length independent, which allows us to process trajectories of arbitrary lengths.

\textbf{Convolutional module --}
The first main convolutional module allows us to expand the trajectory dimension with several convolutional filters. This provides the following layers with a richer embedding based on short-range correlations. 

We build it following the XResNet~\cite{he2019bag} architecture. As we show in~\cref{fig:ml_model}, it consists of an initial convolutional layer, commonly referred to as the \textit{stem}, followed by a series of \textit{residual blocks} that feature a convolutional layer with a skip connection. 
We use one-dimensional convolutions with a kernel size of three and stride one to preserve the trajectory size.  
However, we use a kernel size of one in the skip connections, which can act as the identity or a scaling factor whose main purpose is to match the tensor shapes on both paths of the residual blocks, as we explain below.
This module can take a batch of input trajectories of size $\left[\text{\texttt{batch\_size}}\times T\times d\right]$ and output a batch of features of size $\left[\text{\texttt{batch\_size}}\times T\times\text{\texttt{embedding\_size}}\right]$.

Throughout the architecture, we add a batch normalization layer directly after every convolutional layer, and we use the rectified linear unit (ReLU) activation function by default, except in the last output layer. 

To produce the results in~\cref{sec:results}, we use a single convolutional layer and a ReLU activation in the stem. We use 64 filters to predict the diffusion coefficient and 32 filters for the anomalous diffusion exponent. Then, we have added three residual blocks with 128, 256 and 512 filters, respectively. Hence, $\text{\texttt{embedding\_size}}=512$. In these blocks, the convolutional paths have two convolutional layers: the first one increases the embedding size and the second one preserves the dimensions. In the skip connection, we only have one convolutional layer that increases the embedding size to match the dimensions of the convolutional path. We implement the ReLU activation at the end of the block, after we add the outcome of both paths.

\textbf{Self-attention module --}
We process the features extracted by the convolutional module with a self-attention mechanism that allows the model to capture long-range correlations.

More precisely, we implement a \textit{transformer encoder}, as it was introduced in Ref.~\cite{vaswani2017attention}. 
As we illustrate in~\cref{fig:ml_model}, the encoder block has two main parts, both featuring a skip connection followed by a layer normalization after the sum of both paths. 
In the first one, we have a multi-head attention layer that feeds on the input and, in the second one, we have a couple of pointwise feedforward layers, which are equally applied to each element in the incoming tensor.
Furthermore, we can add a positional encoding before the first encoder block, which provides information about the relative position of each element in the trajectory.
This module can process a batch of embeddings preserving its dimensions. Hence, the input and the output both have size $\left[\text{\texttt{batch\_size}}\times T\times\text{\texttt{embedding\_size}}\right]$.

To produce the results in~\cref{sec:results}, we use four transformer encoder blocks with eight heads in the multi-head attention layers. The pointwise feedforward part adds two fully-connected layers with \texttt{embedding\_size} neurons each, i.e, 512 in this case. Interestingly, we have found that, after the convolutions, the positional encoding has very little impact on the results. Therefore, in the interest of simplicity, we have not used it to obtain the results reported in this work.

\textbf{Feedforward module --}
The last main part is a shallow feedforward fully-connected network that acts element-wise on the features extracted by the previous module. We tailor this part to the specific task at hand to achieve the desired output with the proper dimensions.

For instance, in a regression task, the output dimension is one and we use a scaled sigmoid activation function at the end to define the output range with some margin, e.g., $\log D\in(-3.1, 3.1)$, $\alpha\in(0,2.05)$.
This margin allows the sigmoid to reach the desired values before it saturates.
In a hypothetical case of classification task (as e.g. classifying between diffusion models as done in Ref.~\cite{munoz2021objective}), the final dimension is the number of classes and we use a softmax activation function. Then, we obtain the predictions by choosing the class with the maximum activation value. 
Hence, we can process a feature batch of size $\left[\text{\texttt{batch\_size}}\times T\times\text{\texttt{embedding\_size}}\right]$ and output their predictions with size $\left[\text{\texttt{batch\_size}}\times T\times\text{\texttt{num\_class}}\right]$. In case that $\text{\texttt{num\_class}}> 1$, as in a classification task, we perform an additional post-processing step to obtain an output of size $\left[\text{\texttt{batch\_size}}\times T\times 1\right]$ with the corresponding predictions at each time step.

\subsection{Training procedure}

We follow a standard gradient-based training procedure for both of our models. The only differences between them arise from the training data and how we process it.

The main training loop consists on:
\begin{enumerate}
    \item Predict the values over a batch of training data.
    \item Compute the loss function with respect to the true values. 
    \item Update the model parameters based on the loss' gradient.
\end{enumerate}

We use batches containing 128 trajectories and the L1 loss function, which corresponds to the mean absolute error. Formally, 
\begin{equation}
    \mathcal{L}_{\text{MAE}}(\mathbf{x}) = \frac{1}{n}\sum_{i=1}^{n}\left|y_i - f(\mathbf{x}_i)\right|\,,
    \label{eq:loss}
\end{equation}
where $f(\mathbf{x}_i)$ is the prediction of the $i$-th trajectory in a batch of $n$ samples.

To perform the parameter update, we use an Adam~\cite{adam} optimizer.
We use the fastai~\cite{fastai} library to choose the learning rate with the learning rate finder tool, typically of the order of $10^{-4}$.
Then, we implement a schedule over the training batches both in the learning rate and its momentum, following the one-cycle policy introduced in Ref.~\cite{Smith2019super_convergence}.
We train our models until the performance in the validation set stabilizes, typically between ten to twenty epochs.

To further prevent overfitting and enhance the model generalization capabilities, we use dropout~\cite{Hinton2012dropout} and weight decay~\cite{Hanson1988weight_decay,Krogh1991weight_decay}.
Additionally, we add Gaussian localization noise at different intensities to the trajectories as a form of data augmentation.

\subsection{Training, validating, and testing our models}
\label{app:datasets}

In order to properly evaluate our models, we generate several independent data sets.
We use one to train and validate our models, and we use the others to test them on unseen scenarios.
All the results that we report throughout this work are obtained using the test sets, which we design to evaluate different aspects of our models.

In~\cref{tab:datasets}, we provide the details about the data sets that we use to train, validate and test our models.
These data sets contain simulated trajectories with their corresponding labels at each time step.
We have two main approaches to simulate the trajectories depending on whether we deal with normal or anomalous diffusion. 
Below, we explain how we generate the data for both cases.

\begin{table*}[t]
    \centering
    \begin{tabular}{l c c c c c c c c}
        \textbf{Task} & \textbf{Models} & $D$ &  $\alpha$ & $\sigma_{\text{noise}}$ & \textbf{Traj. length} & \textbf{Segments} & \textbf{Seg. length} & \textbf{Size}\\
        \hline
        \textbf{Train BM} & Brownian motion & $[-3, 3]$ & 1 & $[-6, 2]$ & 200 & $[2, 5]$ & $[10, 190]$ & 100,000\\
        \textbf{Train AnDi} & all anomalous & 1 & $[0.05, 2]$ & 0.1 & 200 & $[2, 5]$  & $[10, 190]$ & 100,064 \\
        \cref{fig:benchmark}A \& B & Brownian motion & $[-3, 3]$ & 1 & 0 & 200 & $[2, 5]$  & $[10, 190]$ & 48,000 \\
         \cref{fig:benchmark}C, ~\cref{fig:pred_vs_true_models} & \multirow{2}{*}{all anomalous} & \multirow{2}{*}{1} & \multirow{2}{*}{$[0.05, 2]$} & \multirow{2}{*}{$[-5,2]$} & \multirow{2}{*}{200} & \multirow{2}{*}{$[2, 5]$} & \multirow{2}{*}{$[10, 190]$} & \multirow{2}{*}{49,994}\\
         \cref{fig:alpha_models}C \& D &&&&& \\
        \cref{fig:benchmark}D,~\cref{fig:alpha_models}A \& B & all anomalous & 1 & $[0.05, 2]$ & $\{0, 0.1\}$ & 200 & $[2, 5]$ & $[10, 190]$ & 50,000 \\
        \cref{fig:benchmark}E \& F & Brownian motion & $[-3, 3]$ & 1 & 0 & 200 & 2 & $[10, 190]$ & 50,000 \\
        \cref{fig:benchmark}G \& H & FBM & 1 & $[0.05, 1.95]$ & 0 & 200 & 2 & $[10, 190]$ & 40,000 \\
        \cref{fig:sbm} & SBM & 1 & $\{0.1, 0.5\}$ & 0 & 200 & 1 & 200 & 6,000 \\
        \cref{fig:exp_attm}B, C \& D & ATTM & (-6.7, 0) & 0.75 & 0 & 200 & $[1, 51]$ & $[1, 200]$ & 10,000 \\
        \cref{fig:exp_attm}E, F \& G & (experiment) &&&& $[200, 2000]$ &&& 755\\
        \cref{fig:exp_integrin} & (experiment) &&&& $[20, 500]$ &&& 4734\\
        \cref{fig:complementary_D}A & Brownian motion & $[-3, 3]$ & 1 & 0 & $[20, 660]$ & $[1, 11]$ & $\{20, 40, 60\}$ & 22,000 \\
        \cref{fig:complementary_D}B & Brownian motion & $[-3, 3]$ & 1 & 0 & $[40, 660]$ & $[2, 11]$ & $\{20, 40, 60\}$ & 20,000 \\
        \cref{fig:complementary_D}C & Brownian motion & $[-3, 3]$ & 1 & $[-6, 0]$ & 200 & $[2, 5]$ & $[10, 190]$ & 48,000 \\
        \hline
    \end{tabular}
    \caption{\textbf{Data set details for all the results reported throughout this paper.}
    The ranges for $D$ and $\sigma_{\text{noise}}$ are in $\log_{10}$ scale, and we take $\alpha$ intervals of 0.05 within the denoted ranges.
    We use $20\%$ of the training data for validation.
    We take two independent sub-samples of a test set with 199,976 trajectories: one for \cref{fig:benchmark}C,~\cref{fig:alpha_models}C \& D,~\cref{fig:pred_vs_true_models} and the other for ~\cref{fig:benchmark}D,~\cref{fig:alpha_models}A \& B.
    In~\cref{fig:benchmark}D, we consider noiseless trajectories, although we add noise with $\sigma_{\text{noise}}=0.1$ to flat CTRW segments that would result in numerical instabilities for the TA-MSD method.
    To generate the noisy trajectories for~\cref{fig:benchmark}C,~\cref{fig:alpha_models}C \& D and~\cref{fig:pred_vs_true_models}, we add 128 random levels of localization noise to each trajectory in the data set, effectively making about $6.4\times10^6$ trajectories.
    \cref{fig:benchmark}C is contained in~\cref{fig:pred_vs_true_models} and, thus, uses the same data.
    In~\cref{fig:sbm}, we evenly split the two values of $\alpha$ among all trajectories.
    In~\cref{fig:exp_attm}B, C \& D we use ATTM trajectories with $\sigma=0.3$ and $\gamma=0.4$ (see \cite{massignan2014nonergodic,manzo2015weak}, do not confuse with $\sigma_{\text{noise}}$).
    We simulate them randomly sampling $D$ and the segment lengths accordingly, and the values here come from analysing the resulting trajectories, which have 18 different segments on average.
    For~\cref{fig:exp_attm}E, F \& G we consider 755 experimental trajectories of the pathogen-recognition receptor DC-SIGN containing from 200 to 2000 frames sampled at $60\,$Hz.
    We obtain all the results in \cref{fig:exp_integrin} from 4734 trajectories of the integrin $\alpha5\beta1$ containing from 20 to 500 frames sampled at $33\,$Hz.
    The data set for~\cref{fig:complementary_D}B is a sub-set of the one from~\cref{fig:complementary_D}A.
    In~\cref{fig:complementary_D}C, we use the same trajectories from~\cref{fig:benchmark}A \& B  and add 128 random levels of localization noise to each of them, effectively making $6.144\times10^6$ noisy trajectories.}
    \label{tab:datasets}
\end{table*}

While there are some differences between how we simulate and label our trajectories for normal and anomalous diffusion, there are several common factors that hold for all of them. For instance, all segments have constant diffusion properties and they are, at least, ten time steps long.

\textbf{Brownian motion --}
We simulate Brownian motion trajectories by taking uncorrelated Gaussian noise as the trajectory displacements.
We control the diffusion coefficient at each time step with the standard deviation of the Gaussian noise, which corresponds to $\sqrt{2D}$.
This way, we can easily generate segments of arbitrary lengths with a constant diffusion coefficient, $D$, along the trajectories.
Finally, we perform the cumulative sum of the displacements to obtain the trajectory coordinates and we subtract the initial position such that they start at the origin.

We consider diffusion coefficients across six orders of magnitude $D\in[10^{-3}, 10^3]$.
However, we take its logarithm as labels for the regression task, such that $y_i\in[-3, 3]$ at every time step.
This greatly simplifies the problem and allows us to keep a consistent performance across all orders of magnitude.

Additionally, we can simulate experimental localization noise by adding Gaussian noise with standard deviation $\sigma_{\text{noise}}$. We use this as a form of data augmentation during training and to study the model's resilience to noise. See~\cref{tab:datasets} for further details.

\textbf{Anomalous diffusion --}
To simulate anomalous diffusion trajectories, we consider the five diffusion models introduced in \cref{app:diffusion} with their respective anomalous diffusion exponent ranges. 
We generate full trajectories for each model following the same procedure detailed in the Supplementary Material from Ref.~\cite{munoz2021objective} and using the library provided by the authors~\cite{andi_dataset}.
Then, in order to obtain heterogeneous trajectories, we split them into segments and combine them together.
We impose the condition that two consecutive segments must differ, at least, either in the diffusion model or the anomalous diffusion exponent.
Finally, we add Gaussian localization noise, with standard deviation $\sigma_{\text{noise}}$. Then, we normalize the resulting displacements by their standard deviation and subtract the initial position to ensure that the trajectory starts at the origin.

Therefore, we have two labels at each time step: the anomalous diffusion exponent and the diffusion model with which the corresponding segment was generated. 
This allows us to use the same data for both a regression task in the anomalous diffusion exponent and a classification task in the diffusion model.
However, in this work, we have mainly focused on the first one.
Furthermore, we balance all the data sets such that there is an even representation of both the anomalous diffusion exponents and diffusion models throughout all the time steps.

\section{Diffusion coefficient prediction}
\label{app:complementary_D}

\begin{figure*}
    \centering
    \includegraphics[width=0.75\textwidth]{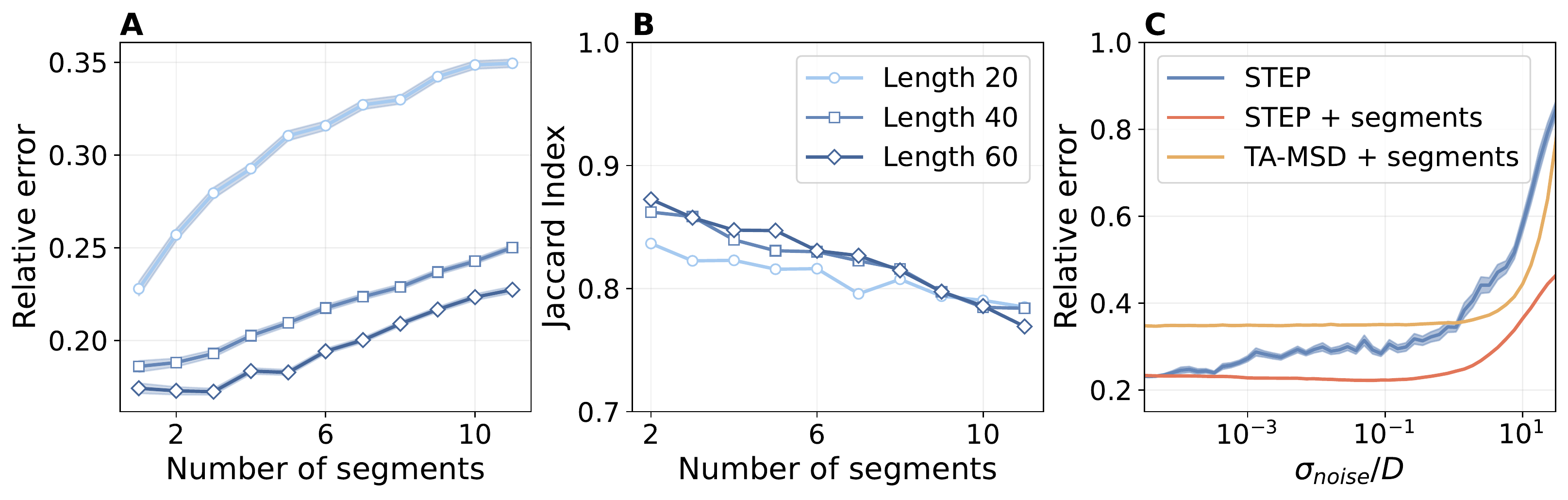}
    \caption{\textbf{\acr{} performance predicting the diffusion coefficient at every time step.} (\textbf{A}) Relative error of \acr{} as a function of the number of segments at three different segment lengths. (\textbf{B}) Prediction accuracy of \acr{} as a function of the number of segments at three different segment lengths. (\textbf{C}) Relative error of \acr{} (blue), \acr{} with known segments (red), and TA-MSD with known segments (yellow) as a function of the ratio between the localization noise's standard deviation $\sigma$ and the diffusion coefficient $D$.}
    \label{fig:complementary_D}
\end{figure*}

In~\cref{sec:results}, we study the capability of \acr{} to properly infer the diffusion coefficient at every time step and detect changes in diffusive behavior.
Here, we complement the analysis presented in the main text by considering additional factors that impact the performance, such as the number of segments in the trajectories and the localization noise, typical of experimental setups.
We show the results in~\cref{fig:complementary_D}.

We investigate the effect of the number of segments on the characterization of the trajectories.
To test it, we fix the segment length and generate trajectories with one to eleven segments (zero to ten changepoints), resulting in trajectories with very different lengths (see~\cref{app:datasets} for details).
In~\cref{fig:complementary_D}A, we see a slight increase of the relative error with the number of segments, although it has a much lesser impact than the segment length, e.g., it is harder to characterize a single segment of twenty points than eight consecutive segments of 40 time steps.
Importantly, even in the presence of 10 changepoints, \acr{} still heavily outperforms the TA-MSD approach applied to segments (no changes) of the same size, e.g., the whole curve for a segment length of 20 in \cref{fig:complementary_D}A is well below the TA-MSD point for a segment length of 20 in \cref{fig:benchmark}C.

In~\cref{fig:benchmark}E and F, we show how to combine \acr{} with a KCPD method to detect diffusion changes.
In \cref{fig:complementary_D}B, we show the performance as a function of the number of segments.
We see that the shortest segments are the hardest to characterize.
However, segment length becomes less important for sufficiently long ones, as the curves for lengths 40 and 60 behave fairly similarly.
We see that \acr{} achieves a better score for shorter segments when the trajectories are very long (11 segments). This suggests that every additional changepoint in the trajectory adds a similar amount of error sources which are eventually outweighed by the accumulated errors along the trajectory as it gets longer.
Nonetheless, even in the most challenging cases with 11 segments, \acr{} correctly detect the vast majority of the points.

Finally, we study the resilience of our method to noise, an important characteristic of experimental trajectories.
In practical scenarios, trajectories are affected by localization noise, which is usually modeled as Gaussian noise of variance $\sigma_{\text{noise}}^2$ added to the trajectories.
Since we consider diffusion coefficients at very different scales along the trajectories, in~\cref{fig:complementary_D}C, we plot the error as a function of the ratio between the noise's standard deviation and the diffusion coefficient.
We see that \acr{} strongly outperforms the TA-MSD approach with known segments even well beyond the noise levels present in relevant experimental scenarios (usually $\sigma_{\text{noise}} / D < 10^{-1}$).
Surprisingly, \acr{} can correctly extract the diffusion coefficient of constant segments (red line) even for enormous errors ($\sigma_{\text{noise}} /D > 10$), showing impressive noise resilience.

\section{Anomalous diffusion exponent prediction for various diffusion models}
\label{app:alpha_by_models}

In \cref{sec:results}, we briefly show how to use \acr{} to study particles that randomly switch between anomalous diffusing states.
Here, we thoroughly characterize the suitability of the method for such task.
We use \acr{} to predict the anomalous diffusion exponent $\alpha$ at every time step of trajectories composed of segments with constant anomalous diffusion exponent and diffusion model, as we detail in~\cref{app:datasets}.

In the main text, we have already studied how the mean absolute error (MAE) in the $\alpha$ prediction depends on the segment length, and we have compared \acr{} to two reference methods.
As an additional reference, we obtain an MAE over all trajectories and time steps of 0.271 with \acr{}, 0.275 with \acr{} and known segments, 0.368 with TA-MSD and known segments, and CONDOR achieved an MAE of 0.237 for trajectories of the same lengths in the AnDi Challenge~\cite{munoz2021objective}.

\begin{figure*}
    \centering
    \includegraphics[width=\textwidth]{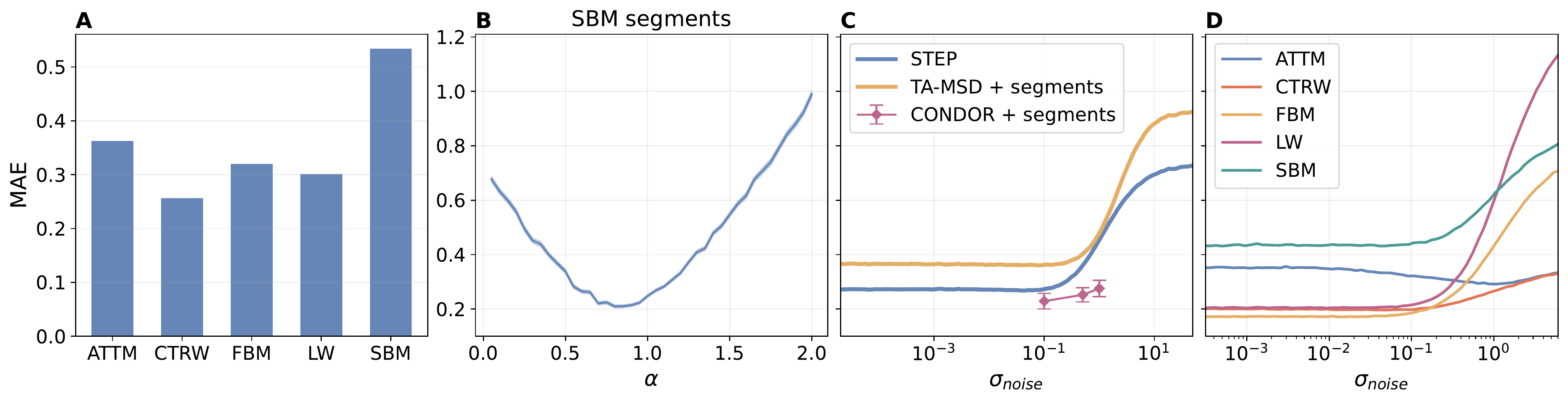}
    \caption{\textbf{\acr{} performance to predict $\alpha$ in terms of the mean absolute error (MAE).} We consider the anomalous diffusion models: ATTM, CTRW, FBM, LW and SBM (see~\cref{app:datasets}). (\textbf{A}) Prediction MAE by diffusion model. (\textbf{B}) MAE as function of $\alpha$ for SBM segments. (\textbf{C}) MAE as a function of the localization noise for \acr{}, TA-MSD, and CONDOR, the latter two with known segments. (\textbf{D}) MAE as a function of the localization noise for \acr{} separated by the diffusion model. The blue line in (\textbf{C}) hence corresponds to the average of the lines presented in this panel.}
    \label{fig:alpha_models}
\end{figure*}

\begin{figure*}
    \centering
    \includegraphics[width=\textwidth]{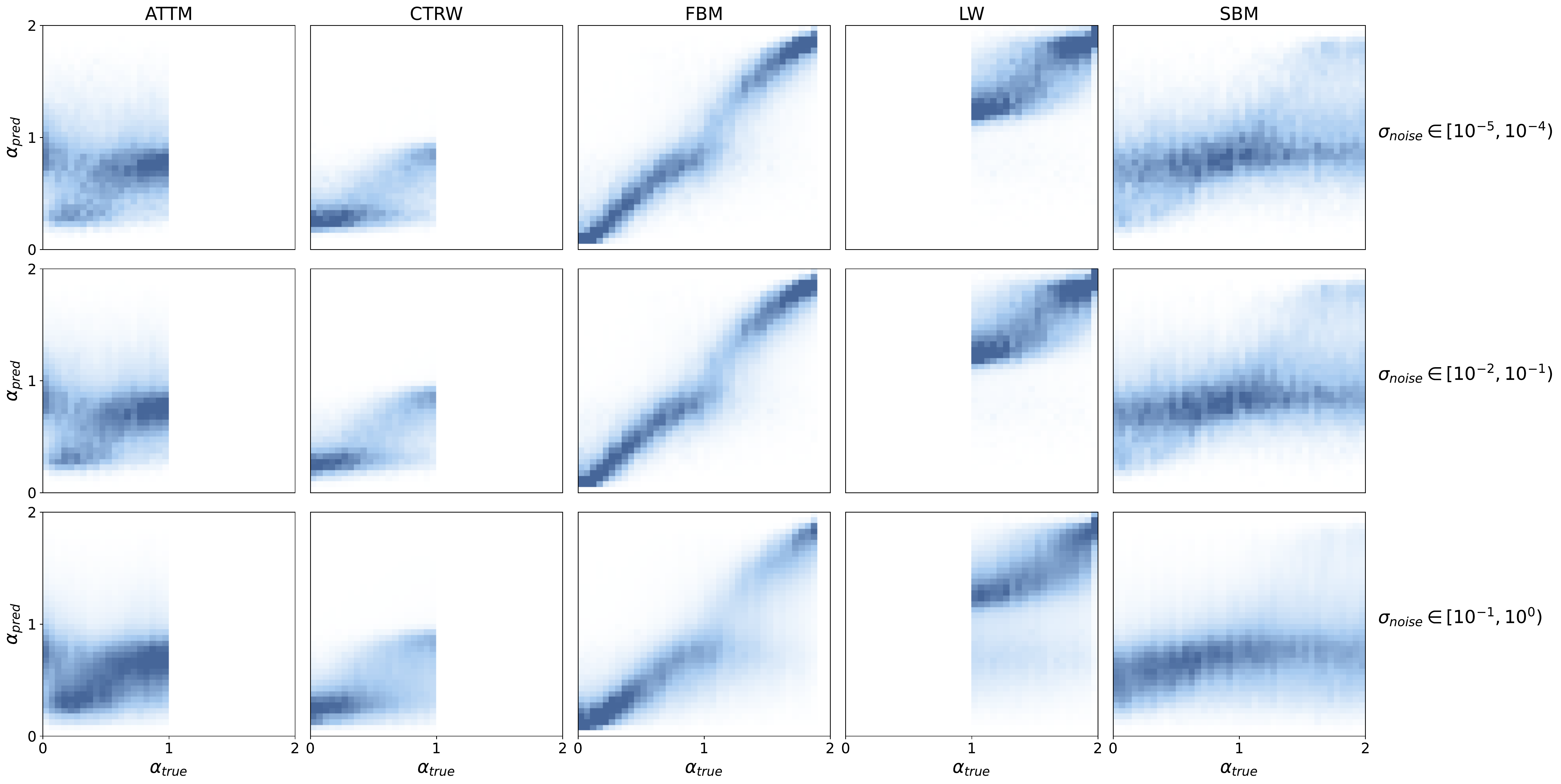}
    \caption{\textbf{Predicted vs true anomalous diffusion exponent.} 2D histograms showing the true and predicted anomalous diffusion exponents for the different diffusion models. Each column contains the histograms belonging to a different diffusion model for three different localization noise levels: $\sigma_{\text{noise}}\in[-5, -4)$ (top), $\sigma_{\text{noise}}\in[-2, -1)$ (middle), and $\sigma_{\text{noise}}\in[-1, 0)$ (bottom). The low-noise FBM histogram (first row, center column) is the same as~\cref{fig:benchmark}C.}
    \label{fig:pred_vs_true_models}
\end{figure*}

Furthermore, we can look at the performance segregated by the diffusion model.
We report the MAE over all segments belonging to each anomalous diffusion model in~\cref{fig:alpha_models}A.
We observe clear differences for some of the models with CTRW, FBM, and LW segments holding the lowest errors.
Additionally, we provide a histogram of predicted and true $\alpha$ for all diffusion models in~\cref{fig:pred_vs_true_models}.

In particular, the MAE in scaled Brownian motion (SBM) segments is significantly larger than in the other models.
This has already been observed in previous works (see for instance Fig. 2d of Ref.~\cite{munoz2021objective}), although the differences here are larger.
A detailed inspection shows that the biggest errors come from shorter segments, in agreement with the results from~\cref{fig:benchmark}D.
This is reasonable since the aging in SBM is the source of the anomalous diffusion~\cite{metzler2014anomalous} (see \cref{sec:continuous_changes}) and therefore it requires longer segments to be correctly characterized.
It is also reasonable to expect the largest errors to happen whenever $\alpha\in(0, 2)$ is close to its range limits and the predictions are in the opposite side.
We see this in in~\cref{fig:alpha_models}B, where the MAE for small and large $\alpha$ are clearly the major contributors to the overall MAE presented in~\cref{fig:alpha_models}A.

\acr{} displays a clear tendency to predict $\alpha\sim0.8$ for SBM segments, as we can see in the right-most column of~\cref{fig:pred_vs_true_models}, which matches the results from~\cref{fig:alpha_models}B.
This behavior is enhanced by the presence of noise, suggesting that the model struggles to identify any clear behavior in short segments, which also happen to be the most common.

Interestingly, we find a similar trend in CTRW segments, where the model has a tendency to predict $\alpha\simeq0.25$, corresponding to nearly immobile particles.
CTRW trajectories are characterized by jumps at random times, resulting in segments in which the particle does not move, usually referred to as waiting times.
Hence, many CTRW segments in our heterogeneous trajectories do not display any movement due to their short lengths, corresponding to a waiting time section.
Therefore, it is impossible for the model to correctly predict $\alpha$, as it does not have any information to work with.

To a lesser extent, we also find that the model predicts $\alpha\sim1$ for low anomalous diffusion exponents in ATTM segments.
In ATTM trajectories with small $\alpha$, we encounter very long segments with low diffusion coefficients.
Similar to the CTRW case, we encounter parts of these long segments in our heterogeneous trajectories containing a unique diffusion coefficient, thus behaving like Brownian motion along the observed time window.
Hence, the predictions $\alpha\sim1$ are correct in these cases.

Finally, we study the resilience of the methods to localization noise, as we do in~\cref{app:complementary_D}.
We present the MAE as a function of the noise standard deviation $\sigma_{\text{noise}}$ in~\cref{fig:alpha_models}C.
We observe a consistent performance of all the methods until reaching considerable levels of noise.
Again, \acr{} is comparable to CONDOR despite the latter having the advantage of knowing the segments beforehand.

As we have seen throughout this section, characterizing some diffusion models is harder than others and the localization noise has a different impact on them, as we show in~\cref{fig:alpha_models}D and~\cref{fig:pred_vs_true_models}.
While increasing the noise level has an overall negative effect, we see that the performance on LW segments suffers the most, while the performance on CTRW segments is barely affected.
Overall, the errors start to increase significantly beyond $\sigma_{\text{noise}}\sim2\times10^{-1}$, which would correspond to harsh experimental conditions.
Interestingly, ATTM segments see a drop in MAE with increasing noise for a limited range.

\end{document}